\documentclass{article}
\usepackage[margin=1in]{geometry}
\usepackage[backend=biber]{biblatex}
\usepackage[dvipsnames]{xcolor}
\usepackage{subcaption}
\usepackage{placeins}   %
\usepackage{sidecap}
\usepackage{booktabs}   %
\usepackage{graphicx}  %
\usepackage{hyperref}  %
\newcounter{subfig}

\usepackage{titlesec}
\titleformat{\subsubsection}[runin]
  {\bfseries}
  {}
  {0pt}
  {}
  [\quad]

\title{Snapshot plots: displaying summary tables as  parallel univariate plots with consistent color highlighting} 
\author{
    Matthias Schonlau \\
    University of Waterloo \\
    \texttt{schonlau@uwaterloo.ca}
    \and
    Sandra Huang  \\
    University of Waterloo  \\
    \texttt{sandra.huang@uwaterloo.ca}
    \and
    Tiancheng Yang  \\
    University of Waterloo  \\
    \texttt{t77yang@uwaterloo.ca}
}

\date{\today}  %

\begin{document}

\maketitle 

\begin{figure}[tbh]
\hspace*{-1cm}
\renewcommand{\thefigure}{1}
    \begin{minipage}[ht]{0.52\linewidth}
        \refstepcounter{subfig}
        \label{f:literacy_highlighting}
        \vspace{-8pt}
        \centering
        \includegraphics[width=\linewidth, clip, trim=78 40 130 60]{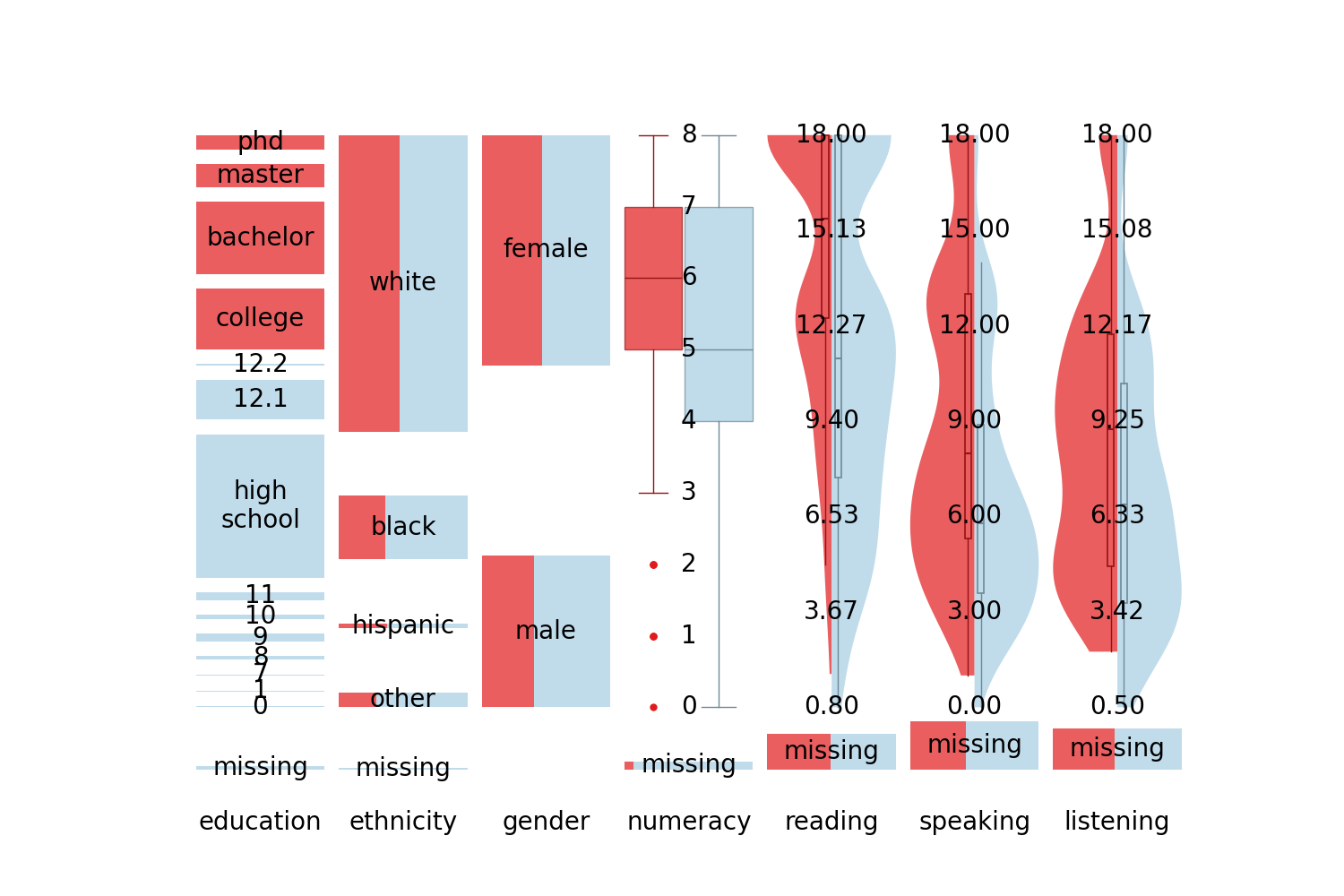}

        \vspace{8pt}
        (a) Snapshot plot 
    \end{minipage}
    \hfill
    \begin{minipage}[ht]{0.48\linewidth}
        \refstepcounter{subfig}
        \label{t:literacy_table1}
        \centering
        {\setlength{\tabcolsep}{3pt}  
            \begin{tabular}{lrrrr}
            \toprule
            \textbf{Variable} & \textbf{Mean} & \textbf{Std.\ Dev.} & \textbf{Min} & \textbf{Max} \\
            \midrule
            \multicolumn{5}{l}{Race / Ethnicity (\%)} \\
            \quad white     & 79.10 & {} & {} & {} \\
            \quad black     & 15.62 & {} & {} & {} \\
            \quad hispanic  & 1.56  & {} & {} & {} \\
            \quad other     & 3.71  & {} & {} & {} \\
            \multicolumn{5}{l}{Gender (\%)} \\
            \quad female    & 60.6 & {} & {} & {} \\
            \quad male      & 39.4 & {} & {} & {} \\
            Education (yrs)       & 13.6 & 2.6 & 0.0  & 21 \\
            \multicolumn{5}{l}{Literacy} \\
            \quad numeracy   & 5.5 & 1.8 & 0.0  & 8  \\
            \quad reading    & 12.7 & 4.7 & 0.8  & 18 \\
            \quad speaking    & 7.8 & 4.5 & 0.0  & 18 \\
            \quad listening   & 8.0 & 4.2 & 0.5  & 18 \\
            \bottomrule
            \end{tabular}
        }
        \vspace{12pt}
        
        (b) ``Table 1''  
    \end{minipage}
\caption{ 
(a) Snapshot plot and (b) the corresponding ``Table 1'' of the literacy study. Here, the snapshot plot highlights participants with college education or higher.  The snapshot plot shows stacked bar charts (with optional spacing) for categorical variables (using side-by-side highlighting) and boxplots/ violin plots for numerical variables.}
\end{figure}

\maketitle
\begin{abstract}
  For empirical studies,  social and health scientists give background characteristics of their sample and summarize them in the famous ``Table~1''.  When treatment/ control groups are present, this table gives summary statistics by group to see whether the background characteristics differ by group.
  We propose snapshot plots --- parallel univariate plots with consistent highlighting --- to visualize such tables. 
  Compared to ``Table~1'', such plots  are designed to facilitate comparisons of background characteristics --- in particular among  groups --- 
and give more detail on numerical variables.
  We provide a web app as well as a python implementation of snapshot plots.
Snapshot plots arise as edge cases of hammock plots (parallel coordinate plots for mixed categorical/ numerical data). We demonstrate the usefulness of snapshot plots for two ``Table~1''s.
\end{abstract}

\section{Introduction}

Empirical studies in the health and social sciences usually contain a  ``Table~1'' 
\cite{fonnes2025table_guide,hayes2019table1, murphy2021praise} describing the demographic composition of the study population.  By contrast,  ``Table 2'' usually contains outcomes \cite{fonnes2025table_guide}. 
A typical ``Table~1'' will contain information on participants gender, age, education, race/ethnicity, income and  additional key variables. 
Such a table typically  contains 5--10 variables, with some tables showing fewer or more variables. 

Figure~\ref{t:literacy_table1} gives an example of a ``Table~1''  
for the literacy data \cite{schonlau2011literacy}. The table shows four dimensions of literacy (\textit{numeracy}, \textit{reading}, \textit{speaking}, and \textit{listening}), as well as some demographic variables (\textit{education}, \textit{gender}, \textit{race/ethnicity} [For ease of writing, non-hispanic white and non-hispanic black have been abbreviated to white and black, respectively]).  
The table shows the percentage distribution of the categorical variables (\textit{gender} and \textit{race}) in the sample as well as mean and standard deviation for the numerical variables (\textit{education} and four dimensions of literacy). 
 
Table~\ref{t:asthma_table1} gives another example from an asthma study \cite{mangione2005measuring}. This table also has two  experimental arms (intervention and control) as well as corresponding p-values. Here, several characteristics vary  significantly by experimental arm. 
{
\setlength{\tabcolsep}{3pt}
\begin{table}[tbp]
  \centering
  \begin{tabular}{lrrrl}
    \toprule
    Variable &\shortstack[r]{Inter-\\vention \\ N=385} & \shortstack[r]{Control \\ N=126} & \shortstack[r]{p \\ Value}& \\
    \midrule
    Age in y, mean (SD) & 8.9 (3.5) & 10.5 (3.4) & $<$.0001 & $^{***}$ \\
    Male gender, \% & 57 & 66 & .08 & $^*$\\
    Asthma severity level, \% & & & &  \\
    \quad Mild intermittent & 64 & 50 & .007& $^{***}$ \\
    \quad Mild persistent & 20 & 24 & .33 & \\
    \quad Moderate/severe persistent & 16 & 26 & .02& $^{**}$\\
    Parent's education, \% & & & \\
    \quad Less than high school & 35 & 29 & .18 & \\
    \quad High school & 34 & 37 & .50 & \\
    \quad More than high school & 31 & 34 & .50 & \\
    Household income, \% & & & \\
    \quad <\$15 000 & 32 & 24 & .10& \\
    \quad \$15 000-<\$30 000 & 40 & 39 &  .82& \\
    \quad $\geq$\$30 000 & 28 & 37 & .06& $^*$\\
    Race/ethnicity, \% & & & \\
    \quad Non-Hispanic white & 19 & 43 & $<$.0001& $^{***}$\\
    \quad Non-Hispanic black & 30 & 23 &  .11& \\
    \quad Hispanic & 29 & 22 & .15& \\
    \quad Other & 22 & 12 & .01& $^{**}$\\
    Insurance type, \% & & & \\
    \quad HMO & 44 & 56 & .02& $^{**}$\\
    \quad PPO-FFS & 47 & 40 &  .17& \\
    \quad No insurance & 9 & 4 & .08& $^*$\\
    Comorbidities, \% & & & \\
    \quad $\geq$1 comorbid condition & 57 & 56 & .88& \\
    \bottomrule
  \end{tabular}
  \caption{\label{t:asthma_table1}``Table~1'' of the asthma study. Characteristics of the 511 survey respondents by group (intervention/ control). Table formatting has been slightly changed for readability. $^*P<.10$; \quad$ ^{**}P<.05$; \quad$^{***}P<.01$    }
\end{table}
}

A ``Table~1'' allows readers to assess generalizability or external validity  \cite{murphy2021praise}.
In a clinical trial, a ``Table~1'' also serves to see whether the randomization worked: If the randomization to the experimental arms worked, the sample characteristics in  each experimental arm should be roughly the same. 

Cleveland observed   
``The power of a graph is its ability to enable one to [\ldots] see patterns and structure not readily revealed by other means of studying the data'' \cite{cleveland1984graphical}.
We therefore seek to visualize “Table~1” and comparable tables. 
We also want to provide a tool for the large number of health and social science researchers who use  ``Table~1''s in their publications. In doing so, we give researchers options to choose between a table and a visualization or to use both. 
 We develop design requirements and then propose  snapshot plots.  Like the table, the proposed plot shows marginal distributions.  We use highlighting to facilitate comparisons by group.    
A web app for the snapshot plot is available at 
\url{https://hammock-plot.streamlit.app/}. The Python source code is publicly available at \url{https://github.com/TianchengY/hammock_plot} and the package can be installed via PyPI (pip install hammock\_plot). The data sets and the  replication files that reproduce the snapshot plots are available in the online material.

The main contributions of this paper are: 
1) designing snapshot plots as a visual alternative to the famous ``Table~1'' in the social and health sciences, 
2) identifying snapshot plots as an edge case to the hammock plot, and 
3) providing a Python implementation and a web app.

\section{\label{s:literature} Related work}
We are not aware of any work that specifically targets the visualization of ``Table 1'' summary tables.  

\subsubsection{Visualizing Tabular Data. }
Many visualization of tabular data visualize every single row:  Lineup \cite{gratzl2013lineup}, Bertifier \cite{perin2014bertifier}, Train delay charts \cite{slingsby2024train}. 
For example, Lineup \cite{gratzl2013lineup} considers tabular data of item rankings across multiple variables.  and encodes the multiple ranks into a stacked bar chart. By contrast, Table~1's are based on tabular data with 100s/1000s of rows and aggregate information is preferred.
Taggle \cite{furmanova2020taggle} visualizes both individual rows and aggregations of rows by a categorical variable. For example, if each observation represents a country, we can aggregate to continents. However, the design is row-centric (e.g. countries in Europe) with aggregation elsewhere (e.g. Africa) giving context.

Other visualizations of tabular data rely on interactivity.  TableLens\cite{rao1994table} visualizes large tables but relies on the focus + context strategy which would not work as a substitute for a ``Table~1'' in printed material. 
Other approaches focus on analyzing tabular data as networks \cite{liu2014ploceus} and hierarchical Tables \cite{li2022hitailor} or  relational database tables \cite{stolte2002polaris} that do not apply to Table~1s.

\subsubsection{Visualizing contingency tables. }
Approaches to visualizing contingency tables include mosaic plots \cite{hartigan1981mosaics, kastellec2007graphs_instead_tables}. 
Cox \cite{cox2004graphingcatdata} turns 2 and 3-way tables into graphs.
Contingency tables do not accommodate continuous variables. Visualization often work best for 2--3 variables.

\subsubsection{Visualizing homogeneous  tables. }
In a homogeneous (two-way) table, all rows have the same structure and all columns measure the same ``kind” of thing. 
Such tables lend themselves to heatmaps \cite{eisen1998cluster} or barcharts \cite{ji2024tables_add_visual}. This arrangement assumes visualizing two categorical variables (for rows and columns) and one numerical variable. It is not meant for more than three variables.
 
\subsubsection{Use of Color. } Most  approaches visualize one column at a time. Color is not used at all \cite{perin2014bertifier} or used in stacked bar charts \cite{gratzl2013lineup, furmanova2020taggle} or used differently for different columns \cite{furmanova2020taggle}. Color is not used to connect columns. (Exception: Taggle \cite{furmanova2020taggle} has an example where one single column only is highlighted by an external second variable (Table 3)).

\section{\label{s:snap}Snapshot plots}

\subsection{\label{s:requirements}Design requirements}
We established the following design requirements to visualize summary tables: 
(R1) Able to visualize enough variables to visualize the typical Table~1. This implies the design must accommodate roughly 5--15 variables.  
(R2) Able to visualize the different types of variables that occur in the typical Table~1. Therefore, the design must accommodate a mixture of categorical, ordered categorical, and quantitative variables.
(R3) Table~1 often involve variables with missing values which are usually imputed or dropped. In both cases it is helpful to know the extent of missingness. The visualization should be able to visualize the amount of missing values.
(R4) Some ``Table~1'' display summary statistics by group (e.g. intervention and control, men and women), allowing for between group comparisons. The visualization should facilitate such a comparison.
(R5)  Because printed venues do not support interaction, the visualization must be effective without interactivity.

We do not aim to visualize measures of uncertainty (such as confidence intervals) as they are not present in a typical ``Table~1''.

\subsection{\label{s:design}Design}
\begin{figure}[tbp]
  \centering
  \mbox{} \hfill
  \includegraphics[width=1\linewidth, clip, trim=155 40 130 100]{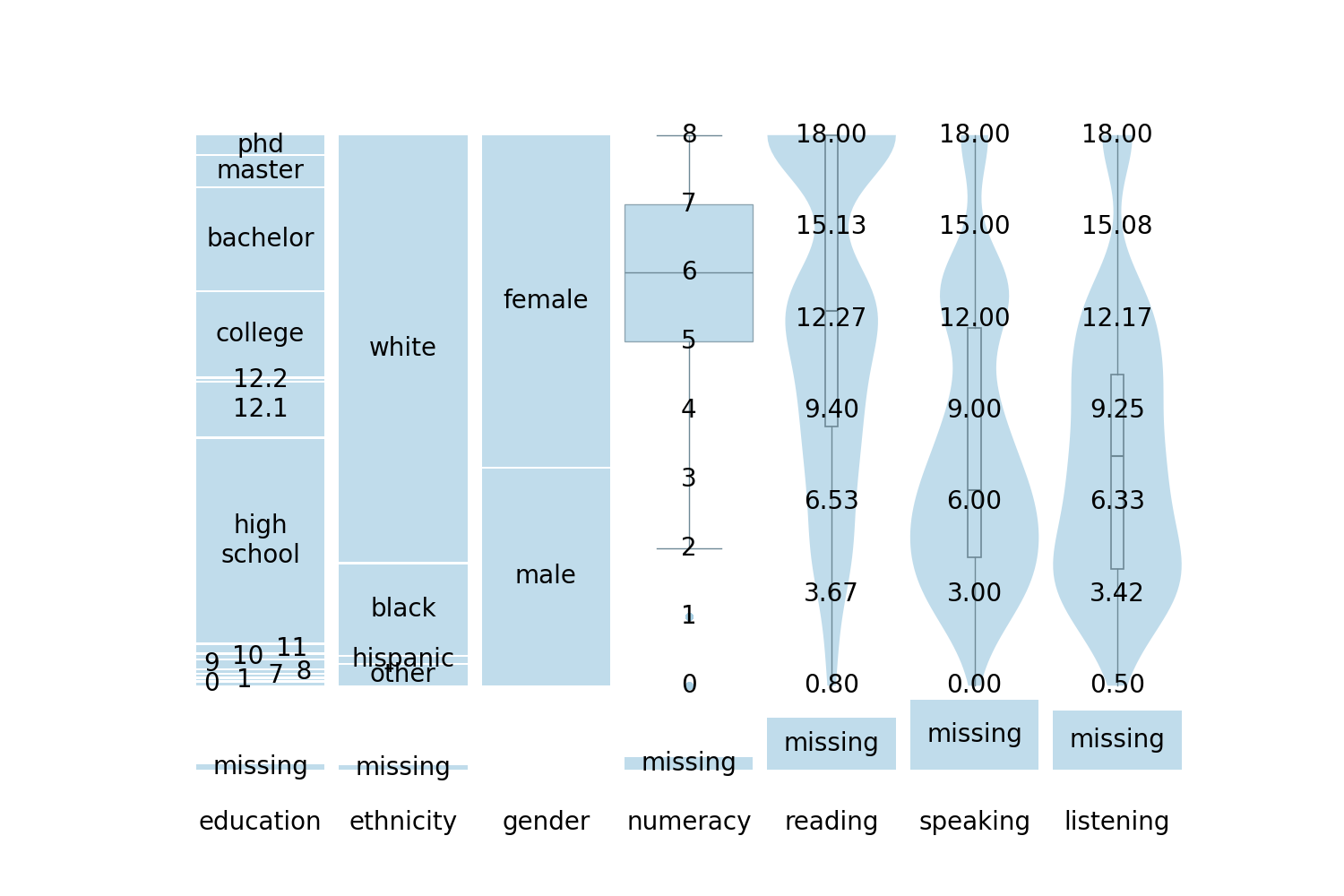}
  \caption{\label{f:literacy_snap_mixed}%
           Snapshot plot of the literacy data with box and  violin plots for numerical variables and stacked barcharts for categorical variables. The extent of missing values is shown at the bottom. For the variable \textit{education}, the values 12.1 and 12.2 refer to more than high school (12 yrs) but less than college. }
\end{figure}

A ``Table~1'' contains univariate (or marginal) statistics. Correspondingly, we propose parallel univariate displays. 
Figure~\ref{f:literacy_snap_mixed} shows a  snapshot plot corresponding to Table~\ref{t:literacy_table1}.
Each variable is shown on a parallel axis, allowing enough space to visualize 5--15 variables (R1). Optionally, we display missing values in a separate row at the bottom of the graph (R3). This design choice can distinguish  missing values from other classes, if desired.
There are different display choices for categorical and numerical variables (R2). These choices as well as highlighting are discussed in the following.

To display numerical variables, we have implemented  two common choices, boxplots and violin plots, and a less common choice we call lumpy rugplots.   

A rugplot shows the distribution of a variable by drawing small tick marks along an axis.  
They appear in scatter plots, density plots,  histograms   \cite{anderson2016histogram_rugplot},  and inside of a bean plot~\cite{kampstra2008beanplot}. 
The tick marks in rugplots are vulnerable to overplotting.
We refer to a lumpy rugplot as a rugplot where the tick marks are replaced with boxes where the width (or height, depending on orientation) of the boxes is proportional to the number of observations it contains. 
Lumpy rugplots may be useful for variables that are inherently numerical but may be subject to overplotting at some values (e.g. counts, where there is heavy overplotting at 0).

To display categorical variables, we implemented a stacked barchart and a horizontal barchart. A horizontal barchart may seem the more obvious choice, but the stacked barchart is advantageous when highlighting and arises naturally within existing visualization systems \cite{hammocksoftware2025}.

\subsubsection{Overplotting for Lumpy Rugplots. }
\begin{figure}[btp]
  \centering
  \mbox{} \hfill
  \includegraphics[width=1\linewidth, clip, trim=155 40 130 100]{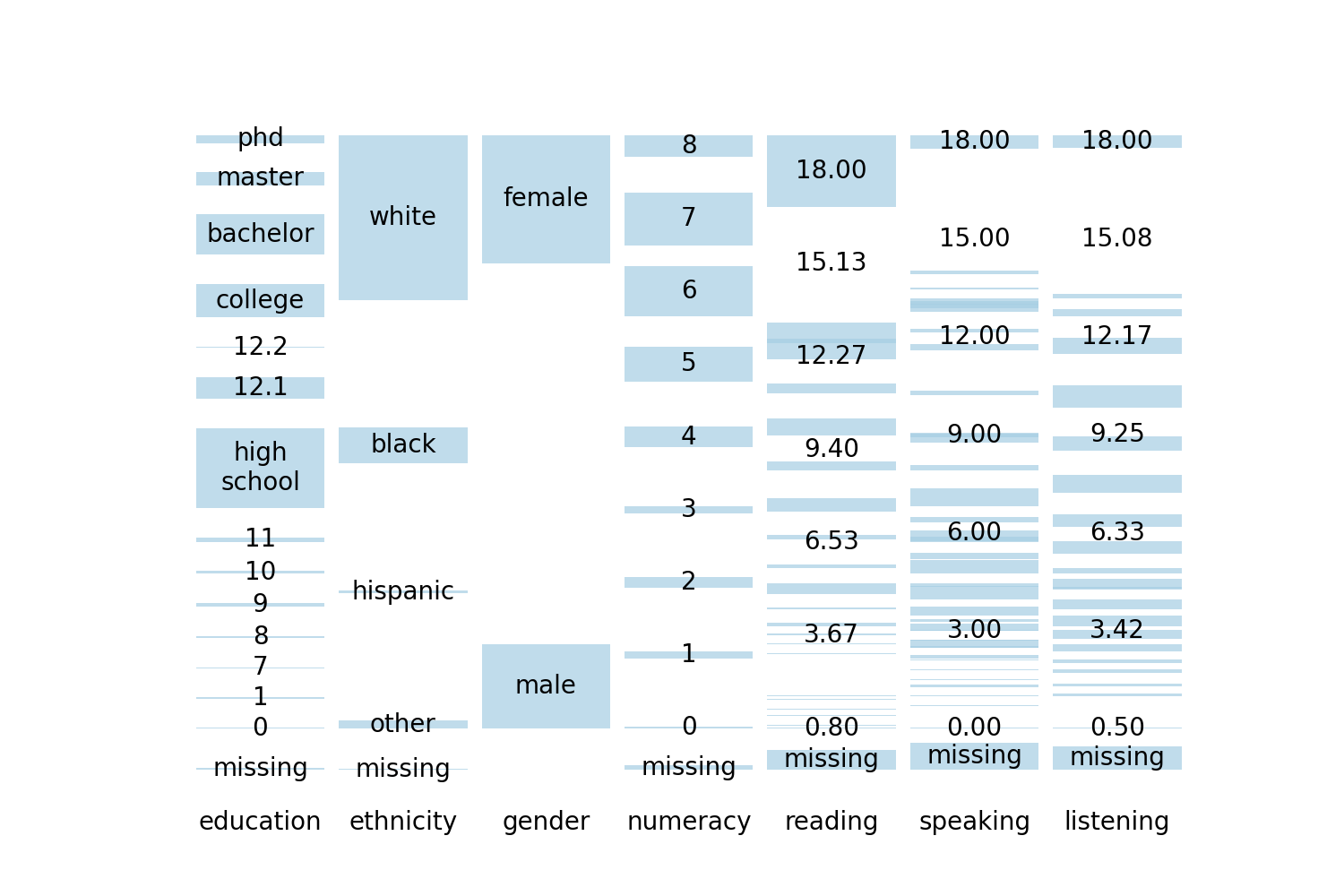}
\caption{\label{f:literacy_snap}%
A snapshot plot of the literacy data.   For numerical variables, this graph shows what we call lumpy rugplots (replacing overplotted tick marks with rectangles). 
The stacked barcharts for \textit{ethnicity} and \textit{gender} contain white space between categories to maintain the ``proportional ink'' principle with the bars in the lumpy rugplots.  
}
\end{figure}

Unlike the violin plot and the boxplot, the lumpy rugplot may be subject to overplotting: adjacent bars may overlap. To prevent this, the bar width of all bars in the lumpy rugplot can be reduced (by reducing the proportionality constant).  
Because of the ``proportional ink'' principle \cite{tufte1983visual}, this has a knock-on effect on the stacked and horizontal barcharts of categorical variables: the proportionality constant must be the same for all variables. 
When the proportionality constant is reduced for the rugplot, and we still want to span the full axis, there must be white space between bars of the stacked barchart.   
Figure~\ref{f:literacy_snap} uses lumpy rugplots for numerical variables and the bar charts with white space in between bars (full-span stacked bar charts). 

Lumpy rugplots and  full-span stacked bar charts look similar except that the former is used for numerical and the latter for  categorical variables. 
This is reminiscent of histograms and barcharts for continuous and categorical variables.  
Even though they have different names, highlighting works identically for both.

\subsubsection{Highlighting. }
When highlighting a category or a numerical range, the corresponding observations in \textit{all} variables are shown in the same color. 
 Figure~\ref{f:literacy_highlighting} highlights all participants with a college degree or more. (Education is measured in years with the following number of years corresponding to degrees: 12 yrs - high school, 14 yrs - 2-year college, 16 - Bachelor, 18 - Master, 21 - Ph.D.)

For violin plots, we exploit the symmetry of the violin plot and create a  half-violin plot for each of two groups. This allows for highlighting one group relative to the remainder (see Figure~\ref{f:literacy_highlighting}).
For boxplots, we display parallel boxplots by group (see Figure~\ref{f:literacy_highlighting}). The width of the boxplot gives a visual cue of group size  \cite{mcgill1978boxplots}. It is possible to highlight multiple groups and complex expressions.

To highlight a categorical variable, we implemented \emph{stack} and \emph{side-by-side}
highlighting.  
 To compare the percentage of highlighted observations across bars of the same axis (column), showing  colours side-by-side (see Figure~\ref{f:literacy_highlighting}) rather than stacked (not shown) is preferable.
Showing colours side-by-side enables comparisons along an aligned axis; showing stacked colors requires comparison of estimated ratios. Facilitating comparisons between  groups meets design requirement (R4).

\subsubsection{Alternative Designs. }
We considered alternatives to visualizing the distribution of numerical variables: {\em Dot plots} \cite{wilkinson1999dotplot} do not scale well with the number of observations. Comparing parallel {\em histograms} is arguably less straight forward than comparing parallel boxplots/ violin plots. 
{\em v-plots} \cite{blumenschein2020vplot} are not specifically designed for parallel comparisons; though they may be a good (if complex) addition in the future.

The snapshot plot displays each variable in a separate column and group(s) are highlighted. Instead, the table layout assigns one column per group. We considered such a layout but found the snapshot layout to be more space efficient.

\section{The asthma study}
\begin{figure}[btp]
  \centering
  \mbox{} \hfill  \includegraphics[width=1\linewidth, clip, trim=155 26 130 100]{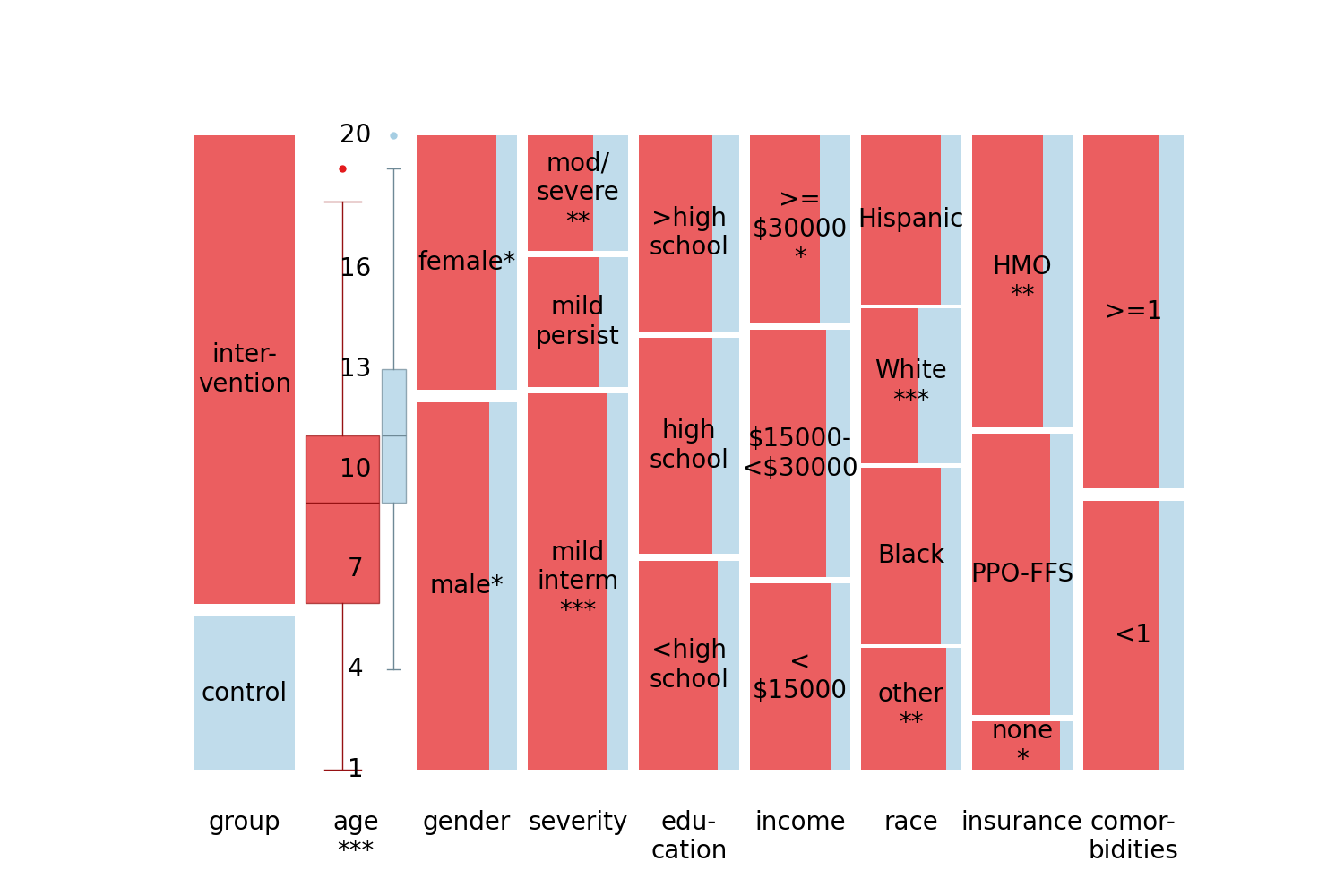}
  \caption{\label{f:asthma_snap}%
Snapshot plot of the sample characteristics for the asthma data. The intervention group is highlighted. As in the table, stars (*) indicate the significance level. 
For example, when considering \textit{race = white}, the 2×2 table (intervention/control × white/all other categories) shows  statistically significant differences.
 }
\end{figure}

The asthma study \cite{mangione2005measuring} evaluated whether an intervention positively influenced health outcomes. 
Health sites (rather than patients) were randomized to intervention or control. The study unit was a patient within a site.
Figure~\ref{f:asthma_snap} shows the snapshot plot corresponding to Table~\ref{t:asthma_table1}.
We see, for example, the patients in the intervention were significantly younger and white patients are underrepresented relative to the control. The stars (***) indicate statistical significance.
(p-values for individual categories were obtained from $\chi^2$ tests on (2,2) tables of intervention/control vs individual category/other categories combined.) 
Because the randomization was at the site level, it is not unusual to find differences at the patient level.

\section{\label{s:discussion}Discussion and conclusion}
Snapshot plots give an overview of tabular data; it is not an row-centric technique.
As such, we see potential benefit during exploratory analysis, as an alternative or an addition to ``Table 1'', giving additional detail on numerical variables that summary tables do not provide.

Snapshot plots arise as edge cases of hammock plots \cite{schonlau2003visualizing,schonlau2024hammock} where bivariate connectors are removed. Hammock plots are plots with parallel coordinates for mixed categorical/numerical data.

To explore scalability, we repeatedly replicated the literacy dataset to create versions with approximately 10,000 and, separately, 1 million observations. 
We then generated Figure \ref{f:literacy_snap_mixed}. Rendering the plot took 0.75 seconds for 10,000 observations and 14.2 seconds for 1 million observations on a laptop. As Table~1's rarely involve more than 10,000 observations, the software successfully accomplishes its intended purpose.

We acknowledge limitations:  
First, the snapshot plot may not be able to accommodate very long tables (landscape mode helps).   
Second, the snapshot plot is intended for summary tables  and  does not accommodate confidence intervals as found in results tables. 
Third, for large numbers of indicator variables, snapshot plots require one column each and are not as effective as tables (one row each).

We introduced snapshot plots, a novel visualization intended for summary tables such as the famous ``Table 1'' in the social and health sciences.  This work focuses on design  rather than empirical evaluation. Future work will evaluate effectiveness of both snapshot plots and lumpy rugplots.

\section*{Acknowledgement}
This work is partially supported  by NSERC grant RGPIN-2025-00392 (Canada).
\printbibliography
\end{document}